# Low-resistance GaN tunnel homojunctions with 150 kA/cm$^2$ current and repeatable negative differential resistance

F. Akyol[1,a)], S. Krishnamoorthy[1], Y. Zhang[1], J. Johnson[2], J. Hwang[2] and S. Rajan[1,2]

[1]Department of Electrical and Computer Engineering, Ohio State University, Columbus, Ohio 43210, USA

[2]Materials Science and Engineering, Ohio State University, Columbus, Ohio 43210, USA

We report GaN n++/p++ interband tunnel junctions with repeatable negative differential resistance and low resistance. Reverse and forward tunneling current densities were observed to increase as Si and Mg doping concentrations were increased. Hysteresis-free, bidirectional negative differential resistance was observed at room temperature from these junctions at a forward voltage of ~1.6-2 V. Thermionic PN junctions with tunnel contact to the p-layer exhibited forward current density of 150 kA/cm$^2$ at 7.6 V, with a low series device resistance of 1 x 10$^{-5}$ ohm.cm$^2$.

[a)] Author to whom correspondence should be addressed.
Electronic mail: akyol.4@osu.edu
Address: 205 Dreese Labs 2015 Neil ave. Columbus, OH, 43210
Tel.: +1-614-688-8458

III-Nitride materials and devices have a broad range of applications, and have already had significant technological and societal impact. In the last two decades, a range of devices have been demonstrated in the III-nitride system including light emitting diodes for solid state lighting[1,2] and communications,[3] lasers for data storage,[4] and high electron mobility transistors for communications [5,6] and power switching .[6,7] More recently, there has been increasing attention to both interband and intra-band[8] tunneling devices. Interband tunneling devices could impact technology by increasing the efficiency of visible LEDs[9,10] and ultraviolet LEDs[11], and by enabling devices for applications such as logic.[12] Negative differential resistance, which has not been observed yet in GaN tunnel junctions, could also enable new device paradigms for logic and high frequency applications.

Interband tunneling in wide band gap materials has intrinsic challenges due to the large band gap that electrons have to surmount increases tunneling probability. Polarization-engineered TJs using GaN/InGaN/GaN hetero-structures were found to decrease the tunneling width and barrier[13-17] and significantly improve tunneling conductance. Recently, homojunction GaN TJs were demonstrated using an n-p-n device structure[18] showing that heavy p- and n-type doping can lead to high tunneling conductance even in wide band gap materials such as GaN. Homojunction GaN TJs were also demonstrated as efficient p-contacts to vertical-cavity surface-emitting lasers[19] and LEDs.[20]

In this work, we investigate the device physics of heavily doped GaN PN tunnel junctions. We find that stand-alone GaN-based homojunctions demonstrate repeatable hysteresis-free room temperature negative differential resistance, and can enable ultra-low resistance contacts for high current density up to 150 kA/cm$^2$.

To understand the operation of GaN PN junctions, two device structures were investigated – stand-alone PN junctions, and NPN structures consisting of a heavily doped tunnel NP junction on top of a lightly doped rectifying conventional PIN junction diode. While stand-alone PN junctions are limited by the resistance of the p-contact, which is typically high, they enable investigation of the intrinsic reverse (Zener) and forward (Esaki) bias characteristics of the PN tunnel junction. NPN structures have low-resistance n-contacts to both terminals of the device and are not limited by contact resistance, and therefore they enable us to probe the intrinsic limit of the reverse (Zener) conductance of the tunnel junction.

Samples were grown using Veeco Gen 930 $N_2$ PAMBE system equipped with elemental sources for Ga, In, Si, and Mg, and a Veeco uni-bulb $N_2$ plasma source with a base pressure of $5 \times 10^{-11}$ Torr. A nominal growth rate of 410 nm/hr was used throughout the growth of all samples under a system pressure of $2.5 \times 10^{-5}$ Torr at a radio-frequency plasma power of 350 W.

Fig. 1(a) shows the epitaxial structure of devices for the stand-alone TJ study. Growth was initiated with 100 nm GaN with Si doping of $4 \times 10^{19}$ cm$^{-3}$ on free standing GaN subtrates (dislocation density ~$3 \times 10^7$ cm$^{-2}$, Si doping ~ $3 \times 10^{18}$, obtained from Saint-Gobain, France). A TJ layer was formed with 10 nm n++ and 20 nm p++ GaN layers. The Si & Mg doping concentrations were designed as $2 \times 10^{20}$ cm$^{-3}$ & $3 \times 10^{20}$ cm$^{-3}$ for sample A, $4 \times 10^{20}$ cm$^{-3}$ & $1 \times 10^{20}$ cm$^{-3}$ for sample B and $4 \times 10^{20}$ cm$^{-3}$ & $5 \times 10^{20}$ cm$^{-3}$ for sample C, respectively. The TJ layer growth was followed by a 180 nm p+GaN and a 20 nm p++GaN layer with Mg doping of $1 \times 10^{20}$ cm$^{-3}$ and $3 \times 10^{20}$ cm$^{-3}$, respectively.

Figure 1 (b) shows the epitaxial structure of an n-p-n diode (sample D) consisting of a TJ in series with a p-i-n diode as depicted in Fig. 1(b). The stack was grown on free-standing GaN substrates and consisted of 100 nm GaN with Si doping of $4 \times 10^{19}$ cm$^{-3}$ layer followed by 15 nm unintentionally doped (uid) GaN and a 100 nm Mg-doped GaN with doping concentration of 6 x $10^{19}$ cm$^{-3}$. Following the p-i-n junction, a TJ structure (same doping concentrations as sample B) was grown and then capped with 150 nm n-GaN (Si=4 x $10^{19}$ cm$^{-3}$), and 20 nm n+GaN (Si=1 x $10^{20}$ cm$^{-3}$) as current spreading and contact layers, respectively. All the Si-doped GaN layers grown at 700 °C under Ga-rich conditions showed streaky reflective high energy electron diffraction (RHEED) patterns, observed at the end of p and n layers as excess metal thermally desorbed from the sample surface, indicating two-dimensional growth. The p-GaN layers were grown under Ga-rich conditions[21] at a growth temperature varying between 610 °C and 680 °C using a Mg beam equivalent pressure of 1 x $10^{-8}$ Torr. During the pGaN growth, a stable In metal film was present on the growth surface to enhance the lateral adatom mobility.[22,23] Both Si and Mg doping concentrations were calibrated with secondary-ion spectroscopy measurements on two separate doping calibration growth stacks (not shown here).

The stand-alone tunnel junction samples (Samples A, B and C) were processed using standard optical lithography techniques. Ni (4 nm) / Au (6 nm) metal layers were deposited using e-beam evaporation on the p-GaN layer and annealed at 450 °C in $O_2$ ambient. Al (20 nm) / Ni (20 nm) / Au (100 nm) metal contacts were deposited on the top n+ GaN layer for Sample D. All the samples were dry etched using BCl3/Cl2 ICP/RIE. The bottom contact to the highly conductive n-GaN substrates was formed by pressing down hand-cleaved In-metal on to the n-GaN substrate.

We analyzed the structural quality of the heaviest doped sample C using atomic force microscopy (AFM) and scanning transmission electron microscopy (STEM). The AFM scan over 5 x 5 µm$^2$ area showed smooth surface morphology with root mean square roughness ~0.5 nm (not shown here). Fig 1 (c) shows STEM Z-contrast images of the sample. No defect features were observed either at the regrowth interface, or at the interface between the n++ GaN and p++ GaN layers. We conclude that no extended defects were induced at the scale of the measurement (~0.17 µm$^2$), with a lower bound on the generated defect density of 5.9 x 10$^8$ cm$^{-2}$. A more detailed investigation of defect generation in heavily doped layers is outside the scope of this paper. While defect generation at the tunnel junction is not critical for applications that involve just a contact to the p-layer, they are critical for device topologies where active layers are introduced *above* the tunnel junction, such as multiple active region LEDs, and solar cells.

Room temperature (RT) I-V characteristics of the sample A, B and C (device area of 70 x 70 µm$^2$) are shown in Fig. 2(a) in semi-log and linear scales. It can be seen that sample C has the lowest reverse current and asymmetric forward and reverse current profile near 0 V, whereas sample B and C showed much higher current at forward and reverse bias. At -0.1 V (+0.1 V), sample B and C showed 130X (360X) and 260X (486X) higher current compared to sample A, respectively. As expected, current density increases with higher Si and Mg doping concentration. From a comparison of sample A and B, we conclude that both forward and reverse tunneling current increases dramatically with increasing Si doping from 2 x 10$^{20}$ cm$^{-3}$ to 4 x 10$^{20}$ cm$^{-3}$ at a similar Mg doping concentration of 3 x 10$^{20}$ cm$^{-3}$. Both sample B and C showed negative differential resistance (NDR) with a peak current density (PCD) in the range of 150 A/cm$^2$ to 350 A/cm$^2$ at a similar forward voltage range from 1.6 V to 2 V at room temperature (RT). The NDR was more pronounced in sample C with peak to valley current ratio (PVCR) of 1.097 compared to sample B (PVCR=1.01) which we attribute to the higher Mg doping in that sample. Fig. 2(b) shows various devices selected randomly from sample C, suggesting robust NDR over the sample. The inset of Fig. 2(b) shows the NDR region of these devices. It can be seen that there are plateaus and double humped structures in NDR region, which is typically observed in the measurement of Esaki diodes[24-27] and attributed to the oscillations in the measurement circuit.[26,27] Such parasitic effects can be fixed by modifying the measurement circuit.[26,27] Multiple scans were performed on the devices (including positive and negative voltage double sweeps). (Fig. 2(c)). The device showed repeatable characteristics at room temperature in both sweep directions in contrast to previous reports of forward tunneling that showed hysteresis and the absence of negative differential resistance in the reverse sweep[28].

Esaki diodes from various semiconductor material systems typically show PCD at a voltage in the range of 0.1 V - 0.4 V.[24-27] The GaN diodes demonstrated here showed PCD at much higher forward voltage (1.6 – 2 V). This is hypothesized to be due to the deep band tail states in heavily Mg-doped (5 x $10^{20}$ cm$^{-3}$) GaN. Using Kane's equation,[29,30] the density of states (DOS) for the light hole (lh) and heavy hole (hh) were calculated assuming an ionized Mg concentration of 1 x $10^{20}$ cm$^{-3}$ (Fig. 3(a)). It should be noted that higher ionization ratio (~50 %) of Mg atoms has been reported for PAMBE grown GaN with heavy (1.2 x $10^{20}$ cm$^{-3}$) Mg-doping.[31] Fig 3(b) and (c) shows the schematics of a TJ in zero and forward bias, respectively. It can be seen that deep band tail states enable forward tunneling at high forward voltage.

The device resistance of the stand-alone TJs was limited by high p-contact resistance especially in the reverse bias regime, based on TLM measurements of the p-contacts. Therefore an n-p-n device with top and bottom n-type ohmic contacts was investigated to estimate the conductivity of the GaN TJs. In the forward bias operation of p-i-n diode, the TJ is reverse biased. Fig. 4 (a) shows the I-V characteristics of the 10x10 μm$^2$ diode in semi-log and linear scale. It can be seen that the device has 8 orders of rectification between -4 V and 4 V. The device showed 3 V of forward voltage at 20 A/cm$^2$ and reached the maximum continuous wave (cw) current density of 150 kA/cm$^2$ at 7.6 V. These values show that tunnel junctions can enable very high current density with low resistive voltage losses. Fig. 4(b) shows the differential resistance of the diode in the forward bias. The differential resistance at current density of 20 A/cm$^2$, 1 kA/cm$^2$ and 150 kA/cm$^2$ were extracted as 9 x 10$^{-3}$ ohm.cm$^2$, 3.1 x 10$^{-4}$ ohm.cm$^2$ and 1 x 10$^{-5}$ ohm.cm$^2$, respectively.

We now discuss the relevance of these results to electronic and optoelectronic device applications. The low tunneling resistance of 1 x 10$^{-5}$ ohm.cm$^2$ is significant for new LED topologies such as p-contact less LEDs, and bipolar cascade LEDs. Unlike previously reported InGaN-based tunnel junctions[13,14,16,17,28] the tunnel junctions reported here do not have a lower band gap InGaN layer that could lead to absorption losses from underlying active regions. Further investigations of the absorption losses in heavily doped regions (such as from band tail states) are needed. Another advantage of homojunction tunnel diodes over heterojunctions is the absence of strain from lattice mismatch, as well as growth challenges related to the growth of InGaN. The current density of 150 kA/cm$^2$ reported here show that tunnel junctions could enable bipolar III-nitride devices such as lasers and superluminescent LEDs to operate at very high current densities with low conduction losses and low charging time for light modulation applications. GaN tunnel junctions are not only limited to emitters with emission energy lower than GaN, but can also be combined with UV laser structures where the optical modes can be confined away from the tunnel junctions to avoid absorption losses. The demonstration of low defect generation and leakage in the NPN structures could also enable new electronic devices that use tunneling contact to the p-type region.

In summary, repeatable RT NDR was observed from PA-MBE grown stand-alone GaN interband tunnel junctions enabled by heavy Si (4 x 10$^{20}$ cm$^{-3}$) and Mg (5 x 10$^{20}$ cm$^{-3}$) doping. The TJ was then integrated with a p-i-n diode as an ohmic contact to p-GaN. The n-p-n test structure has a forward voltage of 3 V at 20 A/cm$^2$ and could be operated at cw current as high as 150 kA/cm$^2$ with a total differential resistance of 1 x 10$^{-5}$ ohm.cm$^2$.

We acknowledge the funding from National Science Foundation (ECCS- 1408416)

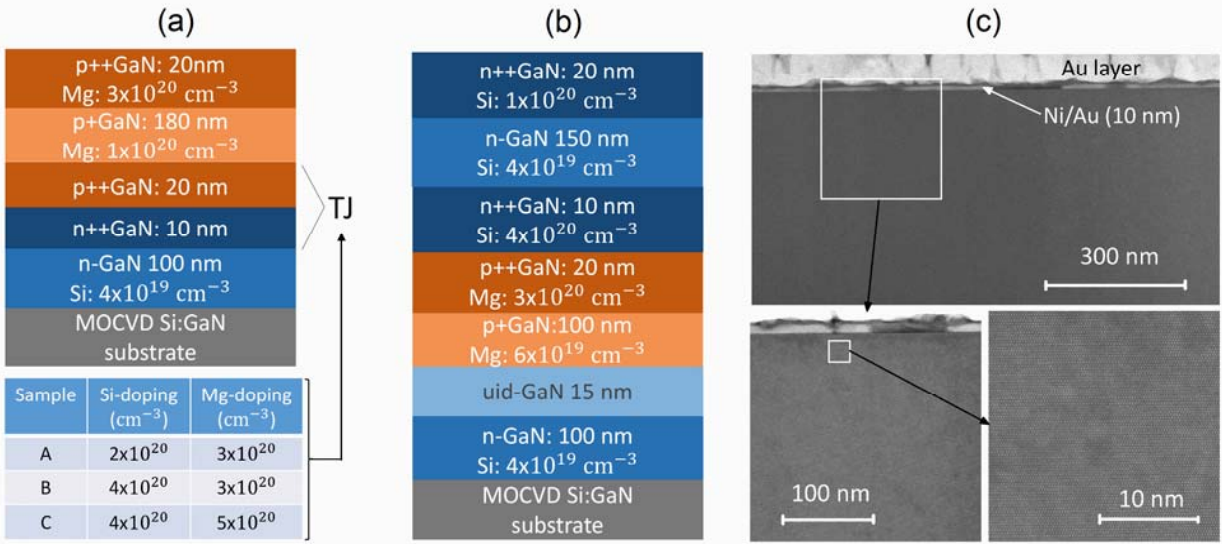

Fig.1. (color online) The epitaxial design of (a) stand-alone TJs of samples A, B, C and (b) n-p-n device (sample D). (c) The STEM Z-contrast cross-sectional images of the sample C showing GaN homojunction device without any indication of nucleation of extended defects.

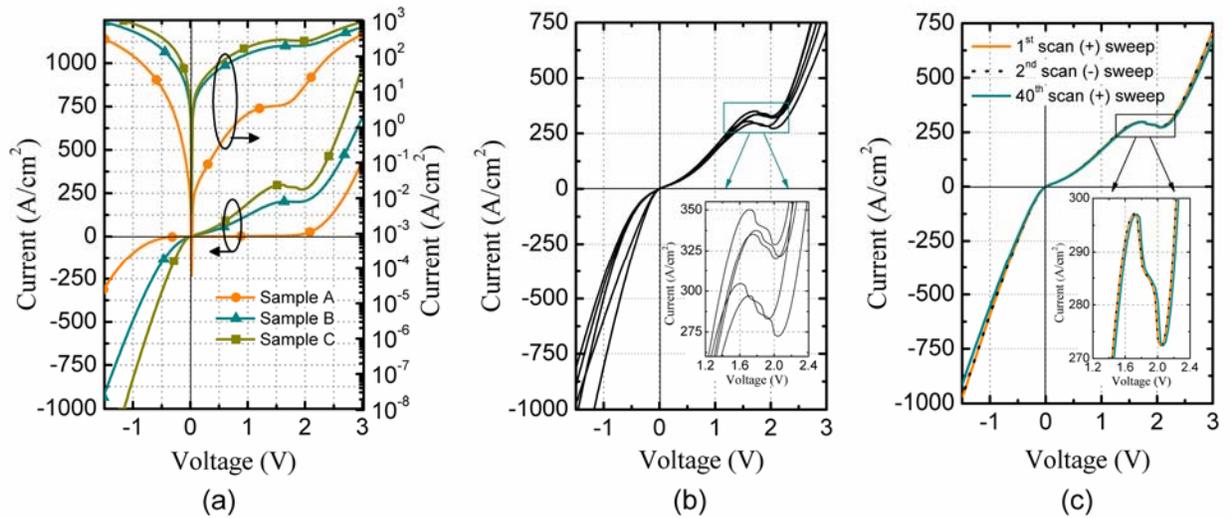

Fig. 2. (color online) Room temperature I-V characteristics of (a) sample A, B, C (device area of 70 x 70 $\mu m^2$) and (b) various devices from sample C showing fairly uniform device characteristics over the sample and (c) a device in sample C with positive (+) and negative (-) voltage sweeps and 40$^{th}$ scan showing repeatable characteristics including the NDR region.

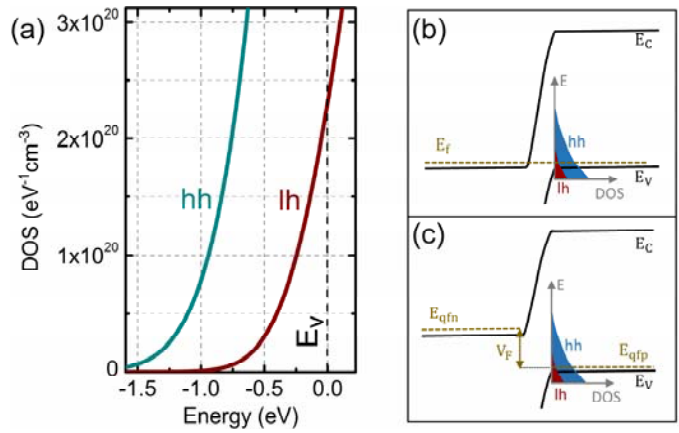

Fig. 3. (color online) (a) Calculated band tail density of states (DOS) of heavy hall (hh) and light hall (lh) bands using Kane's approximation. The schematics of the tunnel junction with tail states under (b) zero and (c) forward bias.

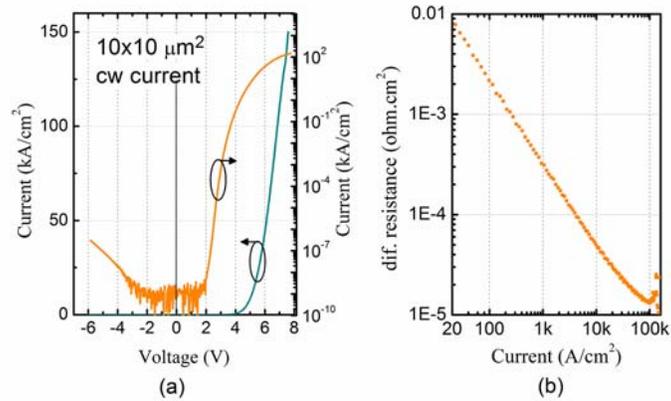

Fig. 4. (color online) (a) The I-V characteristics of sample D in semi-log and linear scale showing 3 V at 20 A/cm$^2$ and a maximum cw current density of 150 kA/cm$^2$. (b) The variation of total differential resistance of the n-p-n diode.

References

1. S. Nakamura, M. Senoh, N. Iwasa, and S. Nagahama, Appl. Phys. Lett. 67, 1868 (1995).

2. T. Mukai, H. Marimatsu, and S. Nakamura, Jpn. J. Appl. Phys. 37, L439 (1998).

3. Martin Dawson, Semiconductor Today magazine, vol 8, 1, Feb 2013.

4. S. Nakamura, and G. Fasol, The blue laser diode, Springer-Verlag, Heidelberg 1997.

5. U. K. Mishra, L. Shen, T. Kazior, and Y. –F. Wu, IEEE Proceedings 96, 2, Feb 2008.

6. Cree Inc, www.cree.com

7. M. Su, C. Chen, and S. Rajan, Semicond. Sci. Technol. 28 (2013) 074012

8. Z. Yang, D. Nath, and S. Rajan, Appl. Phys. Lett. 105, 202111 (2014).

9. F. Akyol, S. Krishnamoorthy, and S. Rajan, Appl. Phys. Lett. 103, 081107 (2013).

10. F. Akyol, S. Krishnamoorthy, Y. Zhang, and S. Rajan, Appl. Phys. Express 8, 082103 (2015).

11. Y. Zhang, S. Krishnamoorthy, J. M. Johnson, F. Akyol, A. Allerman, M. W. Moseley, A. Armstrong, J. Hwang, and S. Rajan, Appl. Phys. Lett. 106, 141103 (2015).

12. W. Li, T. Yu, J. Hoyt, and P. Fay, Device Research Conference p.21, June 2014.

13. S. Krishnamoorthy, D. N. Nath, F. Akyol, P. S. Park, M. Esposto, S. Rajan, Appl. Phys. Lett., 97, 203502 (2010).

14. S. Krishnamoorthy, F. Akyol, P. S. Park, and S. Rajan, Appl. Phys. Lett., 102, 113503 (2013).

15. S. Krishnamoorthy, T. F. Kent, J. Yang, P. S. Park, R. C. Myers, and S. Rajan, Nano Lett., 2013, 13, 2570-2575.

16. Y. Kuwano, M. Kaga, T. Morita, K. Yamashita, K. Yagi, M. Iwaya, Tl Takeuchi, S. Kamiyama, and I. Akasaki, Jpn. J. Appl. Phys. 52 (2013) 08JK12 17  M. Kaga, T. Morita, Y. Kuwano, K. Yamashita, K. Yagi, M. Iwaya, T. Takeuchi, S. Kamiyama, and I. Akasaki, Jpn. J. Appl. Phys. 52 (2013) 08JH06

18  M. Malinverni, D. Martin, and N. Grandjean, Appl. Phys. Lett. 107, 051107 (2015).

19  J. T. Leonard, E. C. Young, B. P. Yonkee, D. A. Cohen, T. Margalith, S. P. DenBaars, J. S. Speck, and S. Nakamura, Appl. Phys. Lett. 107, 091105 (2015).

20  B. P. Yonkee, E. C. Young, J. T. Leonard, S. H. Oh, S. P. DenBaars, J. S. Speck, S. Nakamura, 11th International Conference on Semiconductors, Aug 2015.

21  E. J. Tarsa, B. Heying, X. H. Wu, P. Fini, S. P. DenBaars, and J. S. Speck, J. Appl. Phys. 82, 5472 (1997).

22  Skierbiszewski C, Wasilewski Z, Siekacz M, Feduniewicz A, Pastuszka B, Grzegory I, Leszczynski M and Porowski S 2004 Phys. Status Solidi a 201 320.

23  Neugebauer J, Zywietz T K, Scheffler M, Northrup J E, Chen H and Feenstra R M 2003 Phys. Rev. Lett. 70 56101.

24  J.N. Schulman, and D. H. Chow, IEEE Elect. Dev. Lett. 21, 7, July 2000.

25  M. Oehme, D. Hahnel, J. Werner, M. Kaschel, O. Kirfel, E. Kasper, and J. Schulze, Appl. Ohys. Lett. 95, 242109 (2009).

26  L. Wang, J. M. L. Figueiredo, C. N. Ironside, and E. Wasige, Trans. Elect. Dev. 58, 2 Feb 2011.

27  M. Bao, K. L. Wang, IEEE Trans. Elect. Dev. 53, 10 Oct 2006.

28  S. Krishnamoorthy, P. S. Park and S. Rajan, Appl. Phys. Lett. 99 233504 (2011).

29  P. V. Mieghem Reviews of modern physics 64.3 (1992): 755.

30  E. O. Kane, Physical Review 131.1 (1963): 79.

31 B. Gunning, J. Lowder, M. Moseley, and W. A. Doolittle, Appl. Phys. Lett. 101, 082106 (2012).